\documentclass[aps,twocolumn,floatfix]{revtex4}

\usepackage{amsfonts}
\usepackage{amsmath}
\usepackage{amssymb}
\usepackage{graphicx}
\usepackage{color}
\usepackage{verbatim}
\usepackage{lipsum}
\usepackage{dsfont}
\usepackage[utf8]{inputenc}
\usepackage{physics}
\usepackage{times}

\usepackage[export]{adjustbox}
\usepackage{dcolumn}
\usepackage{bm}
\usepackage[mathlines]{lineno}

\usepackage[T1]{fontenc}

\usepackage{makecell}
\usepackage{upgreek}
\usepackage{amssymb}
\usepackage{tabularx}

\begin{document}

\preprint{APS/123-QED}

\title{

Ultra-fast detection of the center frequency of a spectral line from amplitude-weighted average

}


\author{Ilja Fescenko}
\email{iliafes@gmail.com}

\affiliation{Laser Center, University of Latvia, Latvia}

\date{\today}

\begin{abstract}
Spectroscopy methods often require calculating the central frequency of a resonance line, that is usually implemented by finding a best fit to the spectrum by a line-shape function. Such an iterative procedure is slow and requires an initial guess. We report an analytical method for calculating the central frequency of a spectral line by using the mean value of its frequencies, which are weighted by corresponding normalized intensities. We use this method to calculate two-dimensional arrays of central frequencies from parallely measured magnetic resonance spectra, which are optically detected by a camera sensor in a thin layer of NV centers with superparamagnetic hemozoin crystals on top of it. We demonstrate that our analytical method is more than 800 times faster than the fitting procedure without significant loss of accuracy. For a $400\times400$ pixels sensing array, this method is almost instantaneous, while fitting with the Lorentz function requires several minutes of post-processing on a 4-core MacBook Pro. When a resonance is beyond the spectral range, the amplitude-weighted-mean method does not fail, but it alternates its center frequency proportionally to the mismatch distance. Our method will be useful for spectroscopic applications that require high performance of data processing, allowing their fast optimization, inline processing and real-time output of results. It can also be used for automatic analysis of single spectra, when accuracy of the frequency detection could be slightly ($<10\%$) reduced for sake of stability and simplicity of the data processing. 

\end{abstract}

\maketitle


The precise determination of the frequency of spectral lines underlies many scientific methods such as atomic magnetometry or meteorology. This is usually achieved by finding the best fit to a single-line spectrum using a line-shape function. 
This iterative procedure not only provides information about the center frequency of the fitted spectral line, but also determines its width, amplitude, and baseline. The price to pay for this abundance of information is time-consuming calculations, as well as the need to provide initial guesses for the calculated values.

Although the information provided by the fitting is redundant for many applications, the need for alternative methods has been extinguished in recent decades due to rapid advances in computing technology. For example, the recently widespread NV sensing~\cite{ashfold_nitrogen_2020,barry_sensitivity_2020, mallik_novel_2021} relies on detecting frequencies of magnetic resonances, which are in its DC sensing modalities are simply extracted from the Lorentzian fits~\cite{fescenko_diamond_2019,fescenko_diamond_2020}. In most cases, this is quite justified, since on a moderate personal computer one line is fitted in just a fraction of a second. 

However, the recent increased interest in magnetic imaging~\cite{Fescenko2014,levine_principles_2019,abdelghani_crossover_2023} has revealed the need for an alternative method for determining the center frequency. For example, the processing of the $600\times600$~px$^2$ arrays used in Ref.~\cite{fescenko_diamond_2019} performed $2\times3.6\times10^5$ fitting operations per magnetic image and required more than twenty minutes of computing time. Such fitting could only be done as a post-processing of the accumulated data and proved to be a severe limiting factor for the optimization and development of magnetic imaging techniques. 

In this letter, we propose an alternative method based on the calculation of the mean values of the spectral frequencies, which are weighted by the corresponding values of their intensity amplitudes. We demonstrate the effectiveness and applicability of the new method by applying it to NV magnetic imaging data. Our method shows a performance improvement of more than a few hundred times over the fitting procedure. Moreover, this method is simple, stable and compatible with matrix calculations. This will be of benefit to magnetic imaging and other applications with a similar task of obtaining accurate  positions of spectral lines from their shapes.

\begin{figure*}
      \begin{center}
   \includegraphics[width=0.9\textwidth]{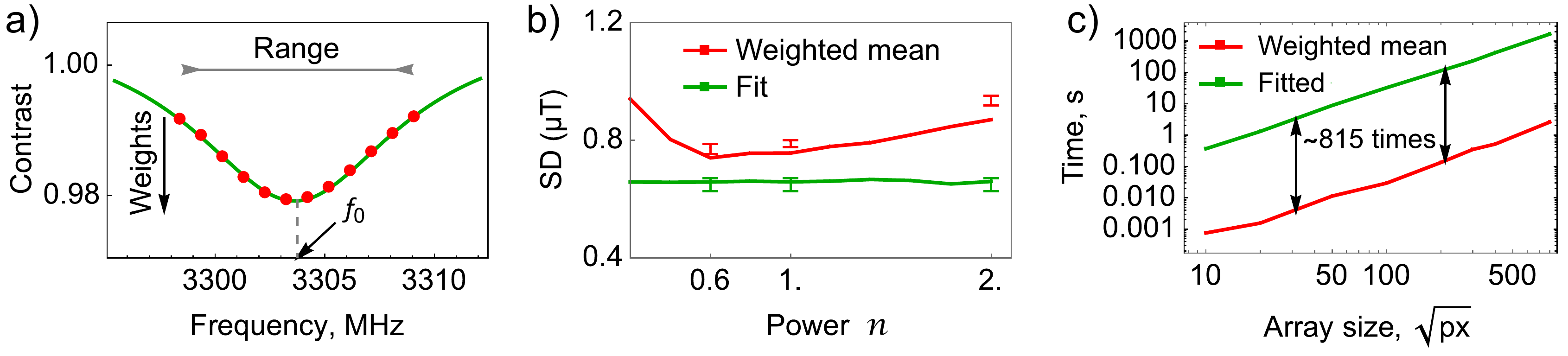}
          \end{center}
       \caption{\textbf{a)} Typical ODMR spectrum: red points are measurements.  \textbf{b)} Spatial standard deviations (SD) of the magnetic field versus weight-scaling power $n$ in the  amplitude-weighted mean calculations (red) compared to SDs obtained from the $n$-independent fitting method (green). Lines are simulation, but points are measurement results. Data ($400\times400$ arrays) are collected during 5 minutes, and additional 7 minutes required for the fitting procedure (0.5 s for the analytic method).
       \textbf{c)} Times for calculation of central frequencies from a dual ODMR experiment (set of $12\times2$ arrays) as a function of array size. The times are given for a 2.5 GHz 4-core MacBook Pro with 16~Gb operative memory. }
  \label{results}
\end{figure*}

The method for calculation of central frequency $f_0$ (see Fig.~\ref{results}a) is based on the weighted arithmetic mean  
\begin{equation}
\overline{f}=\sum_{i=1}^{n} w'_if_i \approx f_0,
 \label{eq:wmean}
\end{equation}
where $f_i$ are in a list of $n$ frequencies in proximity of $f_0$ and the weights $w'_i$ are normalized such that they sum up to 1, i.e., $\sum_{i=1}^{n} w'_i=1$. The weight $w'_i$ is calculated from the signal amplitude at the corresponding frequency $f_i$ in form of signal contrast $C\in(0,1)$ relatively to the fluorescence background, so that the maximum weight falls on the signal peak. The weights are raised to the power of $n=0.6$, which scales the weights for maximum accuracy (see Fig.~\ref{results}b). 
Finally, the weights are normalized by making the following transformation on the original weights:
\begin{equation}
w'_i=\frac{w_i}{\sum_{j=1}^{n} w_j}.
 \label{eq:weigts}
\end{equation}
Unlike the fitting procedure, this analytical method can be directly applied to data matrices, which is especially useful for imaging applications.

To estimate the accuracy, the new method (alongside with the conventional fitting) was first applied to a simple model consisting of a Lorentzian line with added normal noise. The error is defined as the standard deviation (SD) from the given true value $f_0$ of the computed values $f'_0$ over five thousand calculations with random noise. The results of this simulation are shown on Fig.~\ref{results}b as solid lines.

We developed this method to calculate maps of stray magnetic fields from magnetic biocrystals by using optically detected magnetic resonances (ODMR) in NV centers. Such resonances are detected as a dip in fluorescence intensity (see Fig.~\ref{results}a) when polarized ground-state electronic spin sublevels of NV centers are mixed with resonant MW radiation~\cite{barry_sensitivity_2020,levine_principles_2019}. The energy sublevels with non-zero spin projections are split in the magnetic field by Zeeman effect and a measurement of difference of their energies reveals a magnitude of magnetic field along a fixed axis of NV center~\cite{barry_sensitivity_2020,levine_principles_2019}. The further physical and experimental details are described in Refs.~\cite{fescenko_diamond_2019,Berzins_switch_2022}, but only information that is relevant  for performance demonstrations is provided here. 

\begin{figure*}
\begin{center}
\includegraphics[width=0.90\textwidth,valign=t]{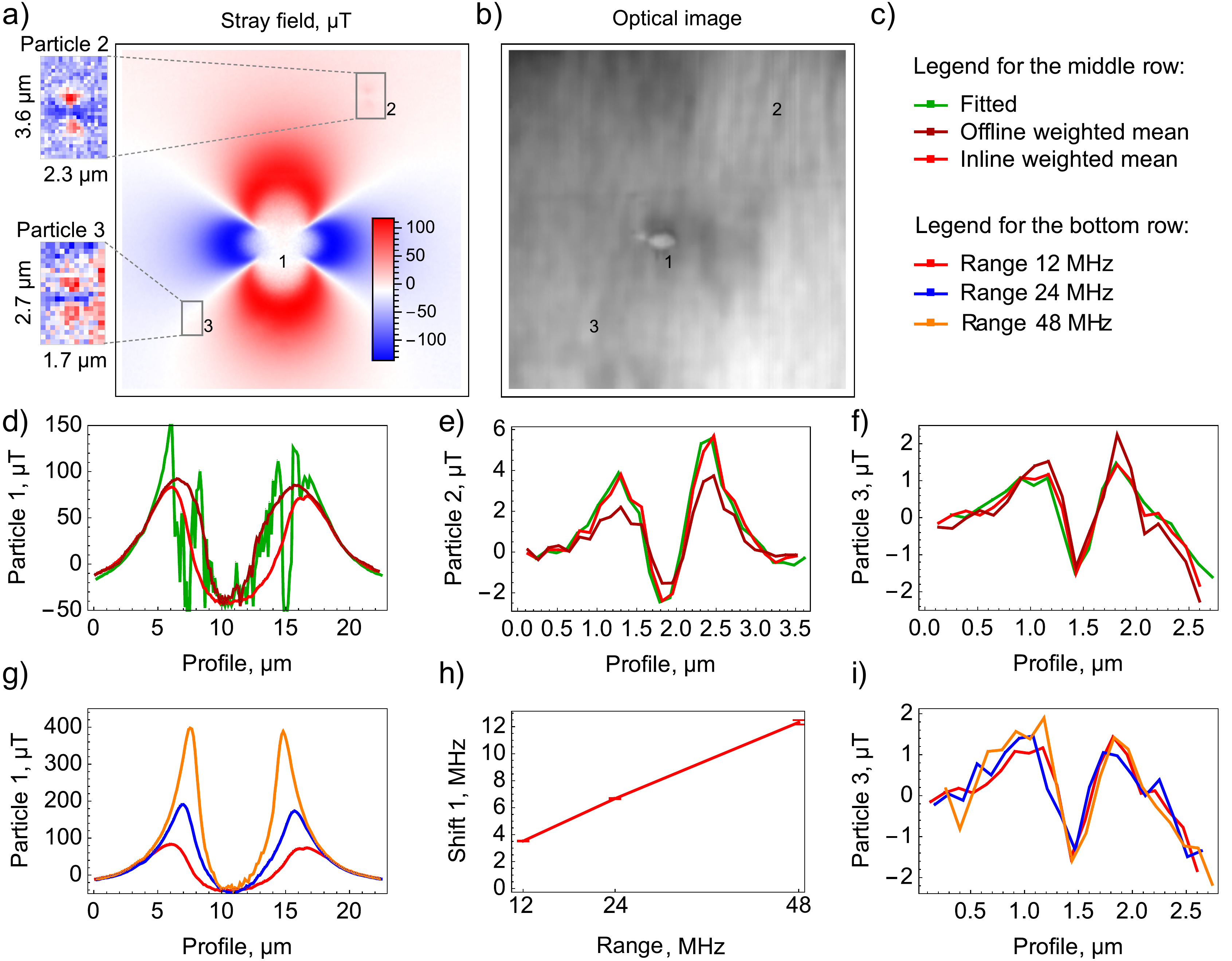}
\hspace{0.1cm}
\end{center}
\caption{\textbf{a)} Obtained by the weighted-mean method magnetic image and \textbf{b)} corresponding optical image of three hemozoin particles on a diamond substrate. \textbf{c)} Legends for d)--g) and i) plots, where "fitted" stays for the conventional fitting by Lorentzian function during post-processing of accumulated raw ODMR data, "offline weighted mean" stays for the post-processing of the same ODMR data with the weighted-mean method, and "inline weighted mean" stays for an immediate processing of the measured ODMR data by the weighted-mean method  with accumulation of the calculated results. "Range" denotes a range of the measured ODMR spectrum (see FIG.~\ref{results}a).  \textbf{d)--f)} Vertical magnetic profiles of three hemozoin particles shown in a).  \textbf{g)} Vertical magnetic profiles of particles 1 obtained by the inline weighted-mean method. \textbf{h)} Magnetic signal (maximum shift from magnetic background) in frequency units detected near particle 1 versus the range of measured spectrum. \textbf{i)} Vertical magnetic profiles of particles 3 obtained by the inline weighted-mean method.
}
\label{Exp}
\end{figure*}

Two ODMR lines are detected in a typical DC modality of NV magnetometry experiment. One of the pair is shown on Fig.~\ref{results}a, where fluorescence intensity is detected in 12 steps by sweeping MW frequency across a magnetic resonance. The value of interest is the central frequency $f_0$, which is proportional to energy difference between mixed spin sublevels. Two resonance are measured because the electronic spin energies could be also shifted by other factors like strain or temperature~\cite{kehayias_imaging_2019,berzins_stress_2021}, while difference between the split energies is due to magnetic field only. Therefore, to obtain a 2D map of magnetic field magnitudes, pairs of central frequencies of ODMR resonances are calculated from their spectra by applying Exp.~\ref{eq:wmean} to matrices of measured fluorescence intensities, and then their differences (or sums, depending on the frequency position relatively ground state level anticrossing) is divided by the NV gyromagnetic ratio. Finally, the magnetizing field $B_{app}$=220~mT is subtracted from the magnetic image to reveal stray magnetic fields originated from magnetized superparamagnetic particles, as shown on Fig. ~\ref{Exp}a. Hemozoin particles (marked by numbers in Fig.~\ref{Exp}b) are studied in detail in Ref.~\cite{fescenko_diamond_2019}. Previously reported superparamagnetic hemozoin crystals had dimensions below optical resolution (like particles 2 and 3 in this letter), whereas micrometer-sized particle 1, which likely generates the huge stray-field signal, was not previously observed. 

We experimentally estimate an accuracy of the weighted-mean method by calculating spatial SD in series of ten plots, $24\times24$~px$^2$, in particle-free area of the magnetic image (Fig. ~\ref{Exp}a). The results of these estimates  are shown on Fig. ~\ref{results}b (scattered points), as a function of scaling power $n$, alongside with the accuracy from the simulation (solid lines). The green line plot and points are obtained as a result of fitting, which is fundamentally independent of the parameter $n$. The similar behavior of the model and experimental data suggests the same nature of the dominant noises in them: the random noise of the ODMR spectrum. Using the optimal power $n=0.6$ does not provide significant noise reduction, so it could be neglected ($n=1$). These SD plots show that the fitting method produces $\sim10\%$ less noise than our weighted-mean method when applied to the same dataset. However, it would be premature to extend this statement to the sensitivity of magnetic measurements, since the fitting requires much more computation time.

The computation times required by both methods versus imaging array size is plotted on Fig.~\ref{results}c. Both line plots are obtained by processing resampled datasets in Wolfram Mathematica on a 2.5 GHz 4-core MacBook Pro with 16 Gb operative memory.  It takes seven minutes to process a data set of $400\times400\times24$, while the weighted-mean method fits in half a second. Regardless of the size of the array, the fitting method is $>800$ slower than the weighted-mean method. Interestingly that about a tenfold speedup in Mathematica is due to the use of matrix calculations instead of parallel lopping.

In the literature, magnetic sensitivity is often defined as magnetic noise over the acquisition time of a signal, which implies that the processing time is negligible compared to the acquisition time. For a magnetic imaging, this is not the case if the fitting procedures are involved. For example, in Figure~\ref{results}b, data is accumulated over five minutes and then fitted over seven minutes (green scattered points). It is clear that if we use these seven additional minutes to continue accumulating data and processing them using the weighted-mean method, then we will achieve less noise. Since the noise decreases as the square root of the accumulation time, in 12 minutes we get SD=0.50~$\upmu$T instead of SD=0.77~$\upmu$T in five minutes of accumulation (left red point). This is $>20\%$ less than the noise level (SD=0.65~$\upmu$T, left green point) obtained in five minutes of data accumulation plus seven minutes of data fitting by the Lorentz function. Although the weighted-mean method is $~10\%$ less accurate than the conventional fitting procedure, it increases the effective magnetic sensitivity by $>20\%$ due to a significant reduction of total measurement time.

Even more benefits can come from the ability to process  spectra "inline" instead of post-processing accumulated spectra. First of all, the accumulated data are reduced from arrays of spectra into one array of single values $f_0$. If we sum the deviations of $\Delta f_0$ from the average center frequency $\bar{f_0}\propto B_{app}$, which in our experiment is proportional to the stray fields, we mitigate the slow magnetic drift in the magnetizing field. This is demonstrated in Fig.~\ref{Exp}d, which shows the profiles of the stray magnetic field from highly magnetized particle 1. The red profile is obtained by processing every 20 accumulated spectra using the weighted-mean method and accumulating $\Delta f_0$ deviations over five minutes of total measurement time. The dark red profile is obtained from the same measurement when the weighted-mean method is applied only once after all spectra have been accumulated. The dark red profile is broadened due to a drift of the laboratory magnetic field during the five minutes of measurements, but the red profile is not affected by the drift.

The inline processing also provides a new opportunity to control the spatial drifts of the sample using a prominent magnetic feature as a marker of the initial position of the sample. This becomes important in unstable environments~\cite{fu_sensitive_2020} or when a long time of data accumulation is required. In previous similar measurements [],  markers were etched into the surface of the diamond to create visible features in the fluorescent images. However, magnetic features are usually more contrast than features in fluorescent images, and costly modification of the sensor surface could be avoided.

Finally, consider the case where the $\Delta f_0$ deviation is large. It corresponds to a spectrum in which the peak is close to the edges of the measurement range, or even shifts out of it.  
Such a signal is produced by particle 1 in Fig.~\ref{Exp}a. While working correctly with weak signals from particles 2 and 3 (Fig.~\ref{Exp} (green plots), the fitting method produces an extremely noisy magnetic signal from particle 1. The weighted-mean method gives a smooth image of stray magnetic fields from particle 1, but its field values close to this particle are underestimated. This is shown in Fig.~\ref{Exp}g where we measured the profiles by scanning extended frequency ranges. We define the shift as the maximum magnetic signal minus the minimum magnetic signal of the profile. Linear nature of the shift vs the spectral range (Fig.~\ref{Exp}h) indicates that the stray fields from particle 1 shift the ODMR signals far beyond the measurement range. In this way, it is possible to obtain a qualitative image of out-of-band magnetic signals, which can be useful when working with unknown magnetic fields.

Weak stray fields from particle 3 (Fig.~\ref{Exp}i) are within the measurement range and do not depend on the choice of range. The simulation shows that the center frequency estimate is performed without bias when the peak is not located in the outer quarters of the spectrum. When this is the case, the fitting method becomes unstable, but the weighted average method gives underestimates.


We have proposed and demonstrated a weighted-mean method for calculating center frequencies from ODMR spectra as an alternative to the redundant, slow, and unstable fitting method. Our method is $10\%$ less accurate, but several hundred times faster, allowing $20\%$ more sensitivity by collecting more data per unit time. The method is almost instantaneous and paves the way for the further development of magnetic imaging with a large field of view.

The author expresses his gratitude for the support in the creation of the diamond magnetic microscope to Dr. A. Berzins and H. Grube. The study was supported by ERAF project 1.1.1.5/20/A/001.

\bibliography{references.bib}

\begin{thebibliography}{12}
\expandafter\ifx\csname natexlab\endcsname\relax\def\natexlab#1{#1}\fi
\expandafter\ifx\csname bibnamefont\endcsname\relax
  \def\bibnamefont#1{#1}\fi
\expandafter\ifx\csname bibfnamefont\endcsname\relax
  \def\bibfnamefont#1{#1}\fi
\expandafter\ifx\csname citenamefont\endcsname\relax
  \def\citenamefont#1{#1}\fi
\expandafter\ifx\csname url\endcsname\relax
  \def\url#1{\texttt{#1}}\fi
\expandafter\ifx\csname urlprefix\endcsname\relax\def\urlprefix{URL }\fi
\providecommand{\bibinfo}[2]{#2}
\providecommand{\eprint}[2][]{\url{#2}}

\bibitem[{\citenamefont{Ashfold et~al.}(2020)\citenamefont{Ashfold, Goss,
  Green, May, Newton, and Peaker}}]{ashfold_nitrogen_2020}
\bibinfo{author}{\bibfnamefont{M.~N.~R.} \bibnamefont{Ashfold}},
  \bibinfo{author}{\bibfnamefont{J.~P.} \bibnamefont{Goss}},
  \bibinfo{author}{\bibfnamefont{B.~L.} \bibnamefont{Green}},
  \bibinfo{author}{\bibfnamefont{P.~W.} \bibnamefont{May}},
  \bibinfo{author}{\bibfnamefont{M.~E.} \bibnamefont{Newton}},
  \bibnamefont{and} \bibinfo{author}{\bibfnamefont{C.~V.}
  \bibnamefont{Peaker}}, \bibinfo{journal}{Chemical Reviews}
  \textbf{\bibinfo{volume}{120}}, \bibinfo{pages}{5745} (\bibinfo{year}{2020}),
  ISSN \bibinfo{issn}{0009-2665, 1520-6890},
  \urlprefix\url{https://pubs.acs.org/doi/10.1021/acs.chemrev.9b00518}.

\bibitem[{\citenamefont{Barry et~al.}(2020)\citenamefont{Barry, Schloss, Bauch,
  Turner, Hart, Pham, and Walsworth}}]{barry_sensitivity_2020}
\bibinfo{author}{\bibfnamefont{J.~F.} \bibnamefont{Barry}},
  \bibinfo{author}{\bibfnamefont{J.~M.} \bibnamefont{Schloss}},
  \bibinfo{author}{\bibfnamefont{E.}~\bibnamefont{Bauch}},
  \bibinfo{author}{\bibfnamefont{M.~J.} \bibnamefont{Turner}},
  \bibinfo{author}{\bibfnamefont{C.~A.} \bibnamefont{Hart}},
  \bibinfo{author}{\bibfnamefont{L.~M.} \bibnamefont{Pham}}, \bibnamefont{and}
  \bibinfo{author}{\bibfnamefont{R.~L.} \bibnamefont{Walsworth}},
  \bibinfo{journal}{Reviews of Modern Physics} \textbf{\bibinfo{volume}{92}},
  \bibinfo{pages}{015004} (\bibinfo{year}{2020}), ISSN
  \bibinfo{issn}{0034-6861, 1539-0756},
  \urlprefix\url{https://link.aps.org/doi/10.1103/RevModPhys.92.015004}.

\bibitem[{\citenamefont{Zheng et~al.}(2021)\citenamefont{Zheng, Wickenbrock,
  Chatzidrosos, Bougas, Leefer, Afach, Jarmola, M.~Acosta, Xu, Z.~Iwata
  et~al.}}]{mallik_novel_2021}
\bibinfo{author}{\bibfnamefont{H.}~\bibnamefont{Zheng}},
  \bibinfo{author}{\bibfnamefont{A.}~\bibnamefont{Wickenbrock}},
  \bibinfo{author}{\bibfnamefont{G.}~\bibnamefont{Chatzidrosos}},
  \bibinfo{author}{\bibfnamefont{L.}~\bibnamefont{Bougas}},
  \bibinfo{author}{\bibfnamefont{N.}~\bibnamefont{Leefer}},
  \bibinfo{author}{\bibfnamefont{S.}~\bibnamefont{Afach}},
  \bibinfo{author}{\bibfnamefont{A.}~\bibnamefont{Jarmola}},
  \bibinfo{author}{\bibfnamefont{V.}~\bibnamefont{M.~Acosta}},
  \bibinfo{author}{\bibfnamefont{J.}~\bibnamefont{Xu}},
  \bibinfo{author}{\bibfnamefont{G.}~\bibnamefont{Z.~Iwata}},
  \bibnamefont{et~al.}, in \emph{\bibinfo{booktitle}{Engineering {Applications}
  of {Diamond}}}, edited by
  \bibinfo{editor}{\bibfnamefont{A.}~\bibnamefont{Mallik}}
  (\bibinfo{publisher}{IntechOpen}, \bibinfo{year}{2021}), ISBN
  \bibinfo{isbn}{978-1-83968-531-6 978-1-83968-532-3},
  \urlprefix\url{https://www.intechopen.com/chapters/74938}.

\bibitem[{\citenamefont{Fescenko et~al.}(2019)\citenamefont{Fescenko, Laraoui,
  Smits, Mosavian, Kehayias, Seto, Bougas, Jarmola, and
  Acosta}}]{fescenko_diamond_2019}
\bibinfo{author}{\bibfnamefont{I.}~\bibnamefont{Fescenko}},
  \bibinfo{author}{\bibfnamefont{A.}~\bibnamefont{Laraoui}},
  \bibinfo{author}{\bibfnamefont{J.}~\bibnamefont{Smits}},
  \bibinfo{author}{\bibfnamefont{N.}~\bibnamefont{Mosavian}},
  \bibinfo{author}{\bibfnamefont{P.}~\bibnamefont{Kehayias}},
  \bibinfo{author}{\bibfnamefont{J.}~\bibnamefont{Seto}},
  \bibinfo{author}{\bibfnamefont{L.}~\bibnamefont{Bougas}},
  \bibinfo{author}{\bibfnamefont{A.}~\bibnamefont{Jarmola}}, \bibnamefont{and}
  \bibinfo{author}{\bibfnamefont{V.~M.} \bibnamefont{Acosta}},
  \bibinfo{journal}{Physical Review Applied} \textbf{\bibinfo{volume}{11}},
  \bibinfo{pages}{034029} (\bibinfo{year}{2019}), \bibinfo{note}{publisher:
  American Physical Society},
  \urlprefix\url{https://link.aps.org/doi/10.1103/PhysRevApplied.11.034029}.

\bibitem[{\citenamefont{Fescenko et~al.}(2020)\citenamefont{Fescenko, Jarmola,
  Savukov, Kehayias, Smits, Damron, Ristoff, Mosavian, and
  Acosta}}]{fescenko_diamond_2020}
\bibinfo{author}{\bibfnamefont{I.}~\bibnamefont{Fescenko}},
  \bibinfo{author}{\bibfnamefont{A.}~\bibnamefont{Jarmola}},
  \bibinfo{author}{\bibfnamefont{I.}~\bibnamefont{Savukov}},
  \bibinfo{author}{\bibfnamefont{P.}~\bibnamefont{Kehayias}},
  \bibinfo{author}{\bibfnamefont{J.}~\bibnamefont{Smits}},
  \bibinfo{author}{\bibfnamefont{J.}~\bibnamefont{Damron}},
  \bibinfo{author}{\bibfnamefont{N.}~\bibnamefont{Ristoff}},
  \bibinfo{author}{\bibfnamefont{N.}~\bibnamefont{Mosavian}}, \bibnamefont{and}
  \bibinfo{author}{\bibfnamefont{V.~M.} \bibnamefont{Acosta}},
  \bibinfo{journal}{Physical Review Research} \textbf{\bibinfo{volume}{2}},
  \bibinfo{pages}{023394} (\bibinfo{year}{2020}), ISSN
  \bibinfo{issn}{2643-1564},
  \urlprefix\url{https://link.aps.org/doi/10.1103/PhysRevResearch.2.023394}.

\bibitem[{\citenamefont{Fescenko and Weis}(2014)}]{Fescenko2014}
\bibinfo{author}{\bibfnamefont{I.}~\bibnamefont{Fescenko}} \bibnamefont{and}
  \bibinfo{author}{\bibfnamefont{A.}~\bibnamefont{Weis}},
  \bibinfo{journal}{Journal of Physics D: Applied Physics}
  \textbf{\bibinfo{volume}{47}}, \bibinfo{pages}{235001}
  (\bibinfo{year}{2014}),
  \urlprefix\url{https://doi.org/10.1088/0022-3727/47/23/235001}.

\bibitem[{\citenamefont{Levine et~al.}(2019)\citenamefont{Levine, Turner,
  Kehayias, Hart, Langellier, Trubko, Glenn, Fu, and
  Walsworth}}]{levine_principles_2019}
\bibinfo{author}{\bibfnamefont{E.~V.} \bibnamefont{Levine}},
  \bibinfo{author}{\bibfnamefont{M.~J.} \bibnamefont{Turner}},
  \bibinfo{author}{\bibfnamefont{P.}~\bibnamefont{Kehayias}},
  \bibinfo{author}{\bibfnamefont{C.~A.} \bibnamefont{Hart}},
  \bibinfo{author}{\bibfnamefont{N.}~\bibnamefont{Langellier}},
  \bibinfo{author}{\bibfnamefont{R.}~\bibnamefont{Trubko}},
  \bibinfo{author}{\bibfnamefont{D.~R.} \bibnamefont{Glenn}},
  \bibinfo{author}{\bibfnamefont{R.~R.} \bibnamefont{Fu}}, \bibnamefont{and}
  \bibinfo{author}{\bibfnamefont{R.~L.} \bibnamefont{Walsworth}},
  \bibinfo{journal}{Nanophotonics} \textbf{\bibinfo{volume}{8}},
  \bibinfo{pages}{1945} (\bibinfo{year}{2019}), ISSN \bibinfo{issn}{2192-8614},
  \urlprefix\url{https://www.degruyter.com/document/doi/10.1515/nanoph-2019-0209/html}.

\bibitem[{\citenamefont{Lamichhane et~al.}(2023)\citenamefont{Lamichhane,
  McElveen, Erickson, Fescenko, Sun, Timalsina, Guo, Liou, Lai, and
  Laraoui}}]{abdelghani_crossover_2023}
\bibinfo{author}{\bibfnamefont{S.}~\bibnamefont{Lamichhane}},
  \bibinfo{author}{\bibfnamefont{K.~A.} \bibnamefont{McElveen}},
  \bibinfo{author}{\bibfnamefont{A.}~\bibnamefont{Erickson}},
  \bibinfo{author}{\bibfnamefont{I.}~\bibnamefont{Fescenko}},
  \bibinfo{author}{\bibfnamefont{S.}~\bibnamefont{Sun}},
  \bibinfo{author}{\bibfnamefont{R.}~\bibnamefont{Timalsina}},
  \bibinfo{author}{\bibfnamefont{Y.}~\bibnamefont{Guo}},
  \bibinfo{author}{\bibfnamefont{S.-H.} \bibnamefont{Liou}},
  \bibinfo{author}{\bibfnamefont{R.~Y.} \bibnamefont{Lai}}, \bibnamefont{and}
  \bibinfo{author}{\bibfnamefont{A.}~\bibnamefont{Laraoui}},
  \bibinfo{journal}{arXiv} p. \bibinfo{pages}{arXiv:2303.09636}
  (\bibinfo{year}{2023}), \urlprefix\url{https://arxiv.org/abs/2303.09636}.

\bibitem[{\citenamefont{Berzins et~al.}(2022)\citenamefont{Berzins, Grube,
  Lazda, Hannig, Smits, and Fescenko}}]{Berzins_switch_2022}
\bibinfo{author}{\bibfnamefont{A.}~\bibnamefont{Berzins}},
  \bibinfo{author}{\bibfnamefont{H.}~\bibnamefont{Grube}},
  \bibinfo{author}{\bibfnamefont{R.}~\bibnamefont{Lazda}},
  \bibinfo{author}{\bibfnamefont{M.~A.} \bibnamefont{Hannig}},
  \bibinfo{author}{\bibfnamefont{J.}~\bibnamefont{Smits}}, \bibnamefont{and}
  \bibinfo{author}{\bibfnamefont{I.}~\bibnamefont{Fescenko}},
  \bibinfo{journal}{Ultramicroscopy} \textbf{\bibinfo{volume}{242}},
  \bibinfo{pages}{113624} (\bibinfo{year}{2022}),
  \urlprefix\url{https://doi.org/10.1016/j.ultramic.2022.113624}.

\bibitem[{\citenamefont{Kehayias et~al.}(2019)\citenamefont{Kehayias, Turner,
  Trubko, Schloss, Hart, Wesson, Glenn, and Walsworth}}]{kehayias_imaging_2019}
\bibinfo{author}{\bibfnamefont{P.}~\bibnamefont{Kehayias}},
  \bibinfo{author}{\bibfnamefont{M.~J.} \bibnamefont{Turner}},
  \bibinfo{author}{\bibfnamefont{R.}~\bibnamefont{Trubko}},
  \bibinfo{author}{\bibfnamefont{J.~M.} \bibnamefont{Schloss}},
  \bibinfo{author}{\bibfnamefont{C.~A.} \bibnamefont{Hart}},
  \bibinfo{author}{\bibfnamefont{M.}~\bibnamefont{Wesson}},
  \bibinfo{author}{\bibfnamefont{D.~R.} \bibnamefont{Glenn}}, \bibnamefont{and}
  \bibinfo{author}{\bibfnamefont{R.~L.} \bibnamefont{Walsworth}},
  \bibinfo{journal}{Physical Review B} \textbf{\bibinfo{volume}{100}},
  \bibinfo{pages}{174103} (\bibinfo{year}{2019}), \bibinfo{note}{publisher:
  American Physical Society},
  \urlprefix\url{https://link.aps.org/doi/10.1103/PhysRevB.100.174103}.

\bibitem[{\citenamefont{Berzins et~al.}(2021)\citenamefont{Berzins, Smits,
  Petruhins, Rimsa, Mozolevskis, Zubkins, and Fescenko}}]{berzins_stress_2021}
\bibinfo{author}{\bibfnamefont{A.}~\bibnamefont{Berzins}},
  \bibinfo{author}{\bibfnamefont{J.}~\bibnamefont{Smits}},
  \bibinfo{author}{\bibfnamefont{A.}~\bibnamefont{Petruhins}},
  \bibinfo{author}{\bibfnamefont{R.}~\bibnamefont{Rimsa}},
  \bibinfo{author}{\bibfnamefont{G.}~\bibnamefont{Mozolevskis}},
  \bibinfo{author}{\bibfnamefont{M.}~\bibnamefont{Zubkins}}, \bibnamefont{and}
  \bibinfo{author}{\bibfnamefont{I.}~\bibnamefont{Fescenko}},
  \bibinfo{journal}{arXiv} p. \bibinfo{pages}{arXiv:2111.12979}
  (\bibinfo{year}{2021}), \urlprefix\url{https://arxiv.org/abs/2111.12979}.

\bibitem[{\citenamefont{Fu et~al.}(2020)\citenamefont{Fu, Iwata, Wickenbrock,
  and Budker}}]{fu_sensitive_2020}
\bibinfo{author}{\bibfnamefont{K.-M.~C.} \bibnamefont{Fu}},
  \bibinfo{author}{\bibfnamefont{G.~Z.} \bibnamefont{Iwata}},
  \bibinfo{author}{\bibfnamefont{A.}~\bibnamefont{Wickenbrock}},
  \bibnamefont{and} \bibinfo{author}{\bibfnamefont{D.}~\bibnamefont{Budker}},
  \bibinfo{journal}{AVS Quantum Science} \textbf{\bibinfo{volume}{2}},
  \bibinfo{pages}{044702} (\bibinfo{year}{2020}), ISSN
  \bibinfo{issn}{2639-0213},
  \urlprefix\url{http://avs.scitation.org/doi/10.1116/5.0025186}.

\end{thebibliography}

\end{document}